\documentclass[aps,twocolumn,floats,superscriptaddress,prd,nofootinbib]{revtex4}
\usepackage{graphicx}
\usepackage{bm}
\usepackage{amssymb}
\usepackage{amsmath}
\usepackage{cancel}
\usepackage{hyperref}
\usepackage{amsmath}

\usepackage{epstopdf}
\usepackage{natbib}
\usepackage{epsfig}

\newcommand{\bma}{\begin{math}}
\newcommand{\ema}{\end{math}}
\newcommand{\beq}{\begin{equation}}
\newcommand{\eeq}{\end{equation}}
\newcommand{\beqa}{\begin{eqnarray}}
\newcommand{\eeqa}{\end{eqnarray}}
\newcommand{\bc}{\begin{center}}
\newcommand{\ec}{\end{center}} 
\newcommand{\bit}{\begin{itemize}}
\newcommand{\eit}{\end{itemize}}

\font\BFd=cmmib10
\font\BFt=cmmib10
\font\BFs=cmmib10 scaled 700
\font\BFss=cmmib10 scaled 500

\def\bbox#1{%
\relax\ifmmode
\mathchoice
{{\hbox{\BFd #1}}}
{{\hbox{\BFt #1}}}
{{\hbox{\BFs #1}}}
{{\hbox{\BFss #1}}}
\else \mbox{#1} \fi }

\newcommand{\MHz}{\mbox{MHz}}

\usepackage{color}

\definecolor{darkgreen}{cmyk}{0.85,0.2,1.00,0.2}

\begin{document}

\title{The Implications of a Pre-reionization 21 cm Absorption Signal for Fuzzy Dark Matter}

\author{Adam Lidz}
\affiliation{Department of Physics \& Astronomy, University of Pennsylvania, 209 South 33rd Street, Philadelphia, Pennsylvania 19104}

\author{Lam Hui}
\affiliation{Center for Theoretical Physics, Department of Physics, Columbia University, New York, NY 10027}

\begin{abstract}
The EDGES experiment recently announced evidence for a broad absorption feature in the sky-averaged radio spectrum around $78  \MHz$, as may result from absorption in the 21 cm line by neutral hydrogen at $z \sim 15-20$. 
If confirmed, one implication is that the spin temperature of the 21 cm line is coupled
to the gas temperature by $z=20$. The known mechanism for accomplishing this is the Wouthuysen-Field effect, whereby Lyman-alpha photons scatter in the intergalactic medium (IGM) and impact the hyperfine level populations. This suggests that early star formation had already produced a copious Lyman-alpha background by $z=20$, and strongly constrains models in which the linear matter power spectrum is suppressed on small-scales, since halo and star formation are delayed in such scenarios.  Here we consider the case that the dark matter consists of ultra-light axions with macroscopic de Broglie wavelengths (fuzzy dark matter, FDM). We assume that star formation tracks halo formation and adopt two simple models from the current literature for
the halo mass function in FDM.  We further suppose that the fraction of halo baryons which form stars is less than a conservative upper limit of $f_\star \leq 0.05$, and that $\sim 10^4$ Lyman-alpha
to Lyman-limit photons are produced per stellar baryon.
We find that the requirement that the 21 cm spin temperature is coupled to the gas temperature by $z=20$ places a lower-limit on the FDM particle mass of $m_a \geq 5 \times 10^{-21} {\rm eV}$. The constraint is insensitive to the precise minimum mass of halos where stars form. As the global 21 cm measurements are refined, the coupling redshift could change and we quantify how the FDM constraint would be modified. A rough translation of the FDM mass bound to a thermal relic warm dark matter (WDM) mass bound is also provided. 
\end{abstract}

\maketitle

\section{Introduction}

The global average redshifted 21 cm signal contains a wealth of information about the thermal and ionization history of intergalactic and pre-intergalactic hydrogen, and is therefore a powerful
probe of early structure formation, the first luminous sources, and the properties of dark matter \citep{Furlanetto:2006jb,Pritchard2012}. Excitingly, the EDGES experiment recently claimed a first, high statistical significance, detection of a 21 cm absorption signal at $z \sim 15-20$ \citep{Bowman:2018yin}. The claimed absorption signal has some puzzling features (see e.g. 
\cite{Barkana:2018lgd,Munoz:2018pzp,Berlin:2018sjs,Ewall-Wice:2018bzf,Mirocha:2018cih,Hill:2018lfx,Venumadhav:2018uwn}; see also earlier work \cite{Munoz:2015bca}): it is surprisingly deep, broad, and flat. Specifically, the depth of the feature is a factor of $\gtrsim 2$ larger
than the maximum depth expected theoretically. This seems to require a mechanism to cool the gas below the temperature expected (in the minimal case that the gas cools adiabatically after decoupling
from the cosmic microwave background (CMB) temperature at $z \sim 150$) or a significant radio background in addition to the CMB at $z \sim 20$ \citep{Ewall-Wice:2018bzf,Feng:2018rje,Sharma:2018agu}, and precious little heat input into the gas at this redshift. In addition, the broad flat-bottomed feature necessitates a delicate balance between subsequent heating and cooling (or between heating and the growing radio background intensity) over $\Delta z \sim 5$. 

In this work we take the EDGES signal at face value, yet do not attempt to explain it fully; instead, we focus on one important implication of the onset redshift ($z \sim 20$) of the absorption feature. In order to observe neutral hydrogen against the radio background, some process is required to break the tendency of the hyperfine level populations to equilibrate with the radio background, otherwise neutral hydrogen will be neither a net absorber nor a net emitter of 21 cm photons. At $z \sim 20$, the gas density throughout most of the universe is too low for collisions to impact the hyperfine level populations, and ultraviolet (UV) photons redshifting into Lyman-series resonances are the key to avoiding equilibration with the radio background. The absorption and subsequent reemission of
 Lyman-alpha (Ly-$\alpha$) photons \cite{Wouthuysen52,Field58} (produced either directly from photons redshifting into the Ly-$\alpha$ resonance or from decay cascades after photons shift into higher-order Lyman-series lines \cite{Hirata:2005mz,Pritchard:2005an}) can cause atoms to swap hyperfine states. Since the optical depth to Ly-$\alpha$ scattering in the neutral intergalactic medium (IGM) at these redshifts is so large \citep{Gunn:1965hd}, each Ly-$\alpha$ photon typically scatters many times
and the resulting energy exchanges bring the radiation into local thermodynamic equilibrium with the kinetic temperature of the gas \cite{Chen:2003gc,Hirata:2005mz}. 
Once the first stars, galaxies, and accreting black holes turn on and emit a sufficient number of UV photons, neutral hydrogen should hence be seen in absorption, provided the gas temperature is indeed cooler than the CMB temperature at this epoch.  
As we review below, an onset redshift of $z \sim 20$
requires on the order of one Ly-$\alpha$ photon for every ten hydrogen atoms throughout the IGM.  In other words, the EDGES result indicates that ``Cosmic Dawn'' was underway by $z \sim 20$ and
requires some minimal level of star formation a mere $180$ million years after the Big Bang. This has interesting implications for our understanding of the first luminous sources.

Among other things, an early onset to the Cosmic Dawn era may be used to constrain the properties of dark matter.  In particular, it limits entire classes of models in which the initial power spectrum of density fluctuations is suppressed on small spatial scales. In such cases, small dark matter halos are absent and star-formation is consequently delayed; this makes the early onset of the EDGES absorption feature difficult to understand. Here, for the most part, we consider the interesting example of fuzzy dark matter (FDM) \cite{Hu:2000ke}, in which the dark matter consists of ultra-light axion-like scalars (of mass on the order of $m_a \sim 10^{-22} {\rm eV}$) with $\sim$ kiloparsec scale de Broglie wavelengths (the precise wavelength depends on mass and velocity). This possibility preserves the large-scale successes of cold dark matter (CDM), is well-motivated by particle physics considerations, can naturally produce a present day matter density $\Omega_m$ of order unity, may help in resolving possible discrepancies between CDM and small-scale observations (although these small-scale discrepancies may owe to the importance of baryonic processes), and has interesting and distinctive observational signatures \cite{Hui:2016ltb}. One observational consequence of FDM is that
the formation of small-mass halos is suppressed \cite{Hu:2000ke}: this feature may be tested most sharply at high redshifts \cite{Hu:2000ke,Bozek:2014uqa}, since {\em only} small mass halos manage to collapse at high redshift in presently
favored CDM cosmological models. Using the onset of Cosmic Dawn as a test of dark matter properties is similar to previous work on using the timing of reionization as a constraint (e.g. \cite{Hu:2000ke,Barkana:2001gr,Bozek:2014uqa}), but the onset redshift of the earlier Cosmic Dawn epoch should provide still sharper limits. A few earlier papers in the literature have considered the closely related question of how warm dark matter (WDM) impacts Cosmic Dawn: first \cite{Sitwell:2013fpa} and \cite{Mesinger:2013nua} modeled the global 21 cm signal and the 21 cm power spectrum in WDM models. Second, as we were finalizing our calculations \cite{Safarzadeh:2018hhg,Schneider:2018xba} appeared; these papers address the implications of the EDGES signal for WDM models, while the latter study also considers FDM. We add to the first two works by exploring the implications of EDGES, to the first three papers by considering FDM, and to all of these studies through our discussion of the Cosmic Dawn models. Our FDM mass constraint can be translated roughly into a WDM mass constraint, which we also provide; these two limits agree with \cite{Safarzadeh:2018hhg,Schneider:2018xba}.

Throughout we adopt the best-fit cosmological parameters from the Planck 2015 analysis (their TT,TE,EE+lowP case) \cite{Ade:2015xua}: $\Omega_m=0.3156$, $\Omega_\Lambda=0.6844$, 
$\Omega_b h^2 = 0.0225$, 
$H_0=67.27\ {\rm km/s/Mpc}$, $\sigma_8=0.831$, and $n_s=0.9645$. We use the Eisenstein \& Hu transfer function \cite{Eisenstein:1997jh} for our CDM models, and modify this suitably for FDM, as detailed below.

\section{The $z \sim 20$ Ly-$\alpha$ Background and Fuzzy Dark Matter}
\label{sec:lya_fuzzy}

Here we briefly review the relevant 21 cm physics (see e.g. \cite{Pritchard2012,Furlanetto:2006jb} for recent reviews), and discuss the background of $z \sim 20$ Ly-$\alpha$ photons implied by the EDGES measurement.
We  then present the main ingredients of our model for the intensity of the Ly-$\alpha$ background (\S~\ref{sec:mod_lya}) and the halo mass function in FDM (\S~\ref{sec:hmf_fdm}). 
In \S~\ref{sec:sfr_param} we describe our fiducial choices for modeling high redshift star formation.

\subsection{The Spin Temperature and the Wouthuysen-Field Effect Coupling}

One key quantity in understanding the redshift evolution of the globally-averaged 21 cm signal is the spin or excitation temperature of
the transition, which describes the relative abundance of atoms in the different hyperfine states. Specifically, the spin temperature is defined
according to:
\beqa
\frac{n_1}{n_0} = 3 {\rm exp}\left(-T_\star/T_s\right),
\label{eq:tspin_def}
\eeqa
where $n_1$ refers to the abundance of hydrogen atoms in the triplet state and $n_0$ is the number of atoms in the lower energy
singlet state.  The quantity $T_\star=0.0681 K$ is defined by the energy splitting between the hyperfine states as $k_B T_\star = h c/\lambda_{21}$, and the
factor of $3$ reflects the statistical degeneracy of the triplet/singlet states. 

As mentioned in the Introduction, for neutral hydrogen to be a net absorber of radio background photons, the spin temperature of the 21 cm line must be cooler than the temperature
of the radio background photons.  Three processes are thought to determine the ratio of atoms in the different hyperfine states and the equivalent spin temperature: first, there
is the absorption/emission of radio background photons; second are collisions with other particles (predominantly hydrogen atoms); and third,  the Wouthuysen-Field effect \cite{Wouthuysen52,Field58}  from UV photons redshifting into Lyman-series resonances. The first process acts to couple the spin temperature of the line to the temperature of the radio background (usually assumed to be 
set by the temperature of the cosmic microwave background (CMB) although other radio photons may possibly be significant \cite{Ewall-Wice:2018bzf,Feng:2018rje}), while the second two effects couple the spin temperature
to the kinetic temperature of the absorbing gas. At $z \sim 20$ the Wouthuysen-Field effect is key to coupling the spin temperature to the gas temperature and allowing neutral hydrogen to be observable against the radio background. 
In statistical equilibrium\footnote{Statistical equilibrium is guaranteed here since the Hubble expansion time is long compared to the other relevant timescales in the problem \cite{Pritchard2012}.} and ignoring collisional coupling, which should be a very good approximation at $z \sim 20$ (see e.g. Eq. 67 of \cite{Furlanetto:2006jb}), the spin temperature may be written as:
\beqa
T_s^{-1} = \frac{T_\gamma^{-1} + x_\alpha T_k^{-1}}{1 + x_\alpha}.
\label{eq:tspin_inv}
\eeqa
The inverse spin temperature is a weighted average of the inverse radio background temperature, $T_\gamma^{-1}$, and the inverse gas kinetic temperature, $T_k^{-1}$
(assuming $T_* \ll T_s , T_\gamma, T_k$). 
Here $x_\alpha$ is
a coupling constant, which is related to the rate at which Ly-$\alpha$ photons scatter off of hydrogen atoms.
More precisely, $x_\alpha$ is defined as
\begin{equation}
x_\alpha \equiv {P_{10} \over A_{10}} {T_* \over T_\gamma} \, ,
\label{eq:xalphadef}
\end{equation}
where $P_{10}$ and $A_{10}$ are
the rates at which a hydrogen atom in the triplet state
makes a transition to the singlet (ground) state 
via the Wouthuysen-Field effect and via spontaneous decay respectively.

The brightness temperature
contrast in redshift-space between a neutral hydrogen cloud  (with neutral fraction $x_{\rm HI}$, overdensity $\delta$, and spin temperature $T_s$) and the radio background is \cite{Madau:1996cs}:
\beqa
\delta T_b \propto x_{\rm HI} (1 + \delta) \left[1 - \frac{T_\gamma}{T_s} \right].
\label{eq:delta_tb}
\eeqa
The EDGES experiment measures the spatial average of this quantity as a function of frequency/redshift. Note that the spin temperature factor in the brightness temperature equation 
(Eq.~\ref{eq:delta_tb}) may be
written as (using Eq.~\ref{eq:tspin_inv}):
\beqa
1 - \frac{T_\gamma}{T_s} = \frac{x_\alpha}{1 + x_\alpha} \left(1 - \frac{T_\gamma}{T_k} \right).
\label{eq:xalpha_demand}
\eeqa
In the limit that $x_\alpha >> 1$, the coupling saturates and $T_s \rightarrow T_k$, while at $x_\alpha=1$ the brightness temperature contrast is half as large
as in the saturated limit. Inspecting Fig. 2 of the EDGES paper \citep{Bowman:2018yin}, which shows their best fit absorption models for different hardware configurations, it appears the 21 cm absorption signal has begun by $z=20$ but is not yet saturated at this redshift. 
We take this to imply that $x_\alpha=1$ at $z=20$, and adopt this as our fiducial assumption in what follows; we subsequently
explore variations around this choice to test how sensitive our conclusions are to it (see Fig.~\ref{fig:zcouple_fdm}).

The coupling constant may be related to the specific number density of photons passing through the Ly-$\alpha$ resonance. Throughout we follow the convention in this field by considering the angle-averaged specific intensity, $J_{\alpha}(z)$, in units of the number of photons per ${\rm cm}^{-2} {\rm s}^{-1} {\rm Hz}^{-1} {\rm str}^{-1}$. 
(All quantities are {\it proper} rather than {\it co-moving} unless otherwise stated.)
Note that this differs from the usual specific intensity by a factor of energy, 
$h \nu_\alpha$. In this case, the coupling coefficient is given by \citep{Pritchard2012}:
\beqa
x_{\alpha} = \frac{16 \pi^2 T_\star e^2 f_\alpha}{27 A_{10} T_\gamma m_e c} S_\alpha J_\alpha.
\label{eq:coupling_co}
\eeqa
Here $f_\alpha=0.4162$ is the oscillator strength of the Ly-$\alpha$ line, $A_{10} = 2.85 \times 10^{-15} {\rm s}^{-1}$ is the Einstein A-coefficient of the 21 cm transition, and the other fundamental
constants have their usual meanings. 
The quantity $S_\alpha$ is an order unity correction factor that takes into account the detailed shape of the radiation spectrum near the Ly-$\alpha$ resonance \cite{Chen:2003gc}.
\footnote{In what follows we set $S_\alpha$ to unity. Assuming that the gas kinetic temperature at $z=20$ is set by adiabatic cooling after decoupling from the CMB at $z=150$, we find
that $S_\alpha=0.78$ at $z=20$ using the fitting formula of \cite{Furlanetto:2006fs,Chuzhoy:2006au}. If the temperature is a factor of $\sim 2$ smaller than this at $z=20$ (as may be preferred by the EDGES data), then $S_\alpha=0.67$.  Note
that accounting for $S_\alpha$ implies that a larger $J_\alpha$ is required to achieve $x_\alpha=1$ and so including this would strengthen our constraints on FDM. Since this is in any case a small effect, and since $S_\alpha$ depends on the uncertain kinetic temperature, we set $S_\alpha=1$ in what follows.} We also assume that the temperature of the radio background is set
by the CMB temperature: if there are in fact additional radio background photons, then a larger Ly-$\alpha$ radiation field is required to achieve coupling and so this possibility would strengthen our FDM constraints. 

It is also convenient to express $J_\alpha$ in terms of the specific intensity equivalent to one Ly-$\alpha$ photon per hydrogen atom \cite{Chen:2003gc} (denoted by $J_0(z)$):
\beqa
J_0(z) = \frac{c n_H(z)}{4 \pi \nu_\alpha}
\label{eq:one_per_atom}
\eeqa
Using Equation \ref{eq:coupling_co} and plugging-in numbers, we can write:
\beqa
x_{\alpha} = \frac{S_\alpha J_\alpha(z)/J_0(z)}{0.069} \left(\frac{1+z}{21}\right)^{2}
\label{eq:thresh_write}
\eeqa
In other words, a specific intensity of $0.069$ Ly-$\alpha$ photons per hydrogen atom is required to achieve
$x_{\alpha}=1$ at $z=20$. We take this as the lower bound on the Ly-$\alpha$ specific intensity required by
EDGES, and use this to constrain models. 

\subsection{Modeling the Ly-$\alpha$ Background Radiation}
\label{sec:mod_lya}

The specific intensity of Ly-$\alpha$ photons is calculated according to \cite{Barkana:2004vb,Hirata:2005mz,Pritchard:2005an,Furlanetto:2006tf}:
\begin{align}
& J_{\alpha}(z) = \frac{c}{4 \pi} (1+z)^2 \sum\limits_{n=2}^{n_{\rm max}} f_{\rm recycle}(n) \nonumber \\
& \times  \int_z^{z_{\rm max}(n)} d z^\prime \frac{1}{H(z^\prime)} \epsilon(\nu^\prime, z^\prime)
\label{eq:jalpha}
\end{align}
In order to understand this equation, it is helpful to first consider photons that are emitted just below (i.e. redward) of the
Ly-$\beta$ frequency. Such photons redshift and propagate freely until they reach the Ly-$\alpha$ resonance, where they
will scatter and contribute to the Ly-$\alpha$ specific intensity $J_\alpha(z)$ and the Wouthuysen-Field effect. 
Next, consider photons that are emitted at a frequency just below that required to induce a $1 s \rightarrow (n+1) p$ transition.
These photons will redshift until they fall into the $1 s \rightarrow n p $ resonance. This, in turn, leads to a rapid decay cascade of lower energy photons, including some Ly-$\alpha$ photons.
Hence the sum over $n$ reflects the contribution to the Ly-$\alpha$ specific intensity from UV photons that redshift into 
the different Lyman-series resonances, while $f_{\rm recycle}(n)$ is the fraction of cascades from the $n$th energy level that
produce Ly-$\alpha$ photons. We use $f_{\rm recycle}$ as calculated in \cite{Pritchard:2005an}, and follow \cite{Barkana:2004vb}
in taking $n_{\rm max}=23$, although our results are insensitive to this choice.
Here $z_{\rm max}(n)$ is the maximum emission redshift for photons that redshift into the $n$th Lyman-series resonance at redshift $z$:
\beqa
1+z_{\rm max}(n) = (1 + z) \frac{1 - (n+1)^{-2}}{1 - n^{-2}},
\label{eq:zmax}
\eeqa 
while the frequency, $\nu^\prime$, of photons emitted at redshift $z^\prime$ and received at $\nu_n$, $z$ is: 
\beqa
\nu^\prime = \nu_n \frac{(1+z^\prime)}{(1+z)}.
\label{eq:nuprime}
\eeqa
Finally, $H(z^\prime)$ is the Hubble parameter and $\epsilon(\nu^\prime, z^\prime)$ is the (specific) co-moving emissivity of UV photons at
frequency $\nu^\prime$ and emission redshift, $z^\prime$. 

In order to calculate the UV emissivity, we assume that it traces the collapse fraction of dark matter halos as \cite{Barkana:2004vb,Hirata:2005mz,Pritchard:2005an}:
\beqa
\epsilon(\nu^\prime, z^\prime) = \epsilon(\nu^\prime) f_\star n_H(z=0) \frac{df_{\rm coll}(> M_{\rm min}, z^\prime)}{dt}.
\label{eq:emiss}
\eeqa
Here we suppose that halos above some minimum mass $M_{\rm min}$ (or equivalently some threshold virial temperature, $T_{\rm vir}$ at $z=20$), are able to host
star formation, and that a fraction $f_\star$ of these baryons are converted into stars.\footnote{Throughout we ignore any impact of FDM on how star formation itself proceeds. In other words,
we assume that FDM influences only the collapse fraction and that it has no effect on, e.g., $f_\star$.} We neglect any dependence of $f_\star$ on halo mass or redshift. We do not account for
the supersonic relative velocity between baryons and dark matter at early times \cite{Tseliakhovich10}, since previous work finds that this has only a minor impact on the global redshifted 21 cm signal \citep{Fialkov:2013jxh}.
The quantity $df_{\rm coll}(> M_{\rm min}, z^\prime)/dt$ denotes
the time derivative of the halo collapse fraction for halos above $M_{\rm min}$. 
In what follows, we vary $M_{\rm min}$ broadly, although our FDM constraints
are insensitive to the precise value of $M_{\rm min}$. As discussed further below (\S~\ref{sec:sfr_param}), our fiducial choice of star formation efficiency is $f_\star=0.05$. The quantity $n_H(z=0)$ is the
present-day abundance of hydrogen atoms (neutral or otherwise).  
The specific emissivity is assumed to follow a power law in frequency, $\epsilon(\nu) \propto \nu^{-\alpha_s - 1}$.
The emissivity normalization is set so that a given total  number of photons (per stellar baryon), $N_\alpha$, is produced between Ly-$\alpha$ and
the Lyman-limit frequency.  We assume the Pop-II type spectrum from Barkana \& Loeb \cite{Barkana:2004vb} which gives $N_\alpha=9690$ and $\alpha_s=-0.14$. Their Pop-III type
spectrum (with $N_\alpha = 4800$ and $\alpha_s=-1.29$) gives a smaller coupling constant ($x_\alpha$, Eq.~\ref{eq:thresh_write}) by a factor of $\sim 2$ and so adopting this would strengthen our constraints.

\subsection{Modeling the Halo Mass Function in FDM}

\label{sec:hmf_fdm}

The next key ingredient in our modeling is the halo mass function in FDM. Unfortunately, full cosmological simulations that incorporate the
effects of quantum pressure on the FDM dynamics are quite challenging. 
Solving the coupled Schr\"odinger and Poisson equations requires resolving the
de Broglie scale even if one is primarily interested in predictions on large scales.
Although considerable progress has recently been made in simulating FDM \cite{Schive:2014dra,Mocz:2017wlg}, the dynamic range in scale required to capture the mass function has not yet been achieved, and so the halo mass function is somewhat uncertain in FDM. (Recently \cite{Veltmaat:2018dfz} 
introduced a hybrid Sch\"odinger/N-body approach that
should be useful in addressing this problem.)

The macroscopic de Broglie wavelength of the FDM particles implies a limit on how tightly the dark matter particles may be confined in potential wells, and so the gravitational growth
of small-scale perturbations is suppressed \cite{Hu:2000ke}. One consequence of this is that it leads to a cut-off in the power spectrum of initial conditions beneath
the axion Jeans scale at matter-radiation equality \cite{Hu:2000ke}. Quantitatively, the co-moving scale at which the linear power spectrum is reduced by a factor of two (relative to CDM) for an FDM particle mass, $m_a$, is $k_{1/2} = 26\ {\rm Mpc}^{-1} \left[m_a/(5 \times 10^{-21} {\rm eV})\right]^{4/9}$ and the corresponding mass scale is \cite{Hui:2016ltb}:
\begin{align}
M_{1/2} & =  \frac{4 \pi \rho_M}{3} \left(\frac{\pi}{k_{1/2}}\right)^3 \nonumber \\
& =  3.2 \times 10^8 M_\odot \left(\frac{\Omega_m}{0.32}\right) \left(\frac{5 \times 10^{-21} {\rm eV}}{m_a}\right)^{4/3}.
\label{eq:mhalf}
\end{align}
In addition to the suppression of the initial power spectrum of fluctuations, the growth of perturbations on small scales is reduced, and halo formation 
at $z \sim 20$ should be entirely truncated on mass scales below the axion Jeans mass at that redshift. This scale
is much smaller in mass than $M_{1/2}$: for example, with our cosmological parameters and $m_a = 5 \times 10^{-21} {\rm eV}$, $M_J(z=20) = 4.2 \times 10^5 M_\odot$  \cite{Hui:2016ltb}.

The precise form of the halo mass function suppression between $M_{1/2}$ and $M_J$ awaits improved FDM simulations. In order to assess the impact of current uncertainties in modeling
the FDM halo mass function, we consider two different models for the FDM halo mass function. The first one is the fitting formula of Schive et al. \cite{Schive:2015kza} (specifically, their Eq. 7): this comes from simulations that include the cut-off in the initial power spectrum (Eq.~\ref{eq:mhalf}), yet ignore the subsequent impact of quantum pressure on the simulation dynamics (i.e. N-body simulations, as opposed to wave simulations, were used).
This is expected to be a good approximation on mass-scales sufficiently larger than the $z=20$ axion Jeans mass. Since the suppression or cut-off scale in the initial conditions ($M_{1/2}$) is on a larger scale than Jeans scale, one expects the main impact of FDM on the halo mass function to come from the initial conditions and not from the fuzzy/wave dynamics.
The Schive et al. \cite{Schive:2015kza} fitting formula gives a simple remapping between the CDM and FDM halo mass functions; we implement this using the CDM mass function according to Sheth \& Tormen \citep{Sheth:1999mn}. 

We also explore the approach of Marsh \& Silk \cite{Marsh:2013ywa}, which aims to semi-analytically model
the mass function, including the suppression between $M_{1/2}$ and $M_J$.  The starting point for their work is to note that -- in contrast to the case of LCDM where the linear growth factor is scale-independent -- small-scale perturbation growth is suppressed in FDM.  They further suppose that the impact on the halo mass function may be modeled by introducing a physically motivated mass-dependent collapse threshold for halo formation, plausibly
describing the suppression between $M_{1/2}$ and
$M_J$. In their model, the halo mass function follows that of Sheth-Tormen \citep{Sheth:1999mn} except that the collapse threshold is modified by a mass-dependent factor, $\mathcal{G}(M)$ with $\nu = \delta_c(z)/\sigma(M) \rightarrow \delta_c(z) \mathcal{G}(M)/\sigma(M)$. Here $\delta_c(z) = 1.686/D(z)$ is the usual threshold linear density for spherical collapse, $D(z)$ is the LCDM linear growth-factor (normalized to $1$ at $z=0$), and $\sigma^2(M)$ is the variance of the linear density field smoothed on mass-scale $M$. The functional form of the mass dependent factor is motivated by considering the scale-dependent growth of linear perturbations in FDM, with the mass calculated as that enclosed in a sphere of radius $R = \pi/k$ at the cosmic mean density.
In practice, we use the fitting formulae in Eqs. 15-19 of  \cite{Marsh:2016vgj} to describe $\mathcal{G}(M)$\footnote{These formulae provide convenient fits to the full mass-dependent factor, $\mathcal{G}(M)$, calculated in Marsh \& Silk \cite{Marsh:2013ywa}, and so we use these fitting formulae in what follows. We refer to this as the ``Marsh \&  Silk mass function''.} and the FDM transfer function from Eq. 3 of that reference to compute $\sigma(M)$ in FDM. 
We note that the Marsh \& Silk \citep{Marsh:2013ywa} mass function shows a strong suppression below $M \sim 0.01 M_{1/2}$. 

\begin{figure}[htpb]
\bc
\includegraphics[width=1.0\columnwidth]{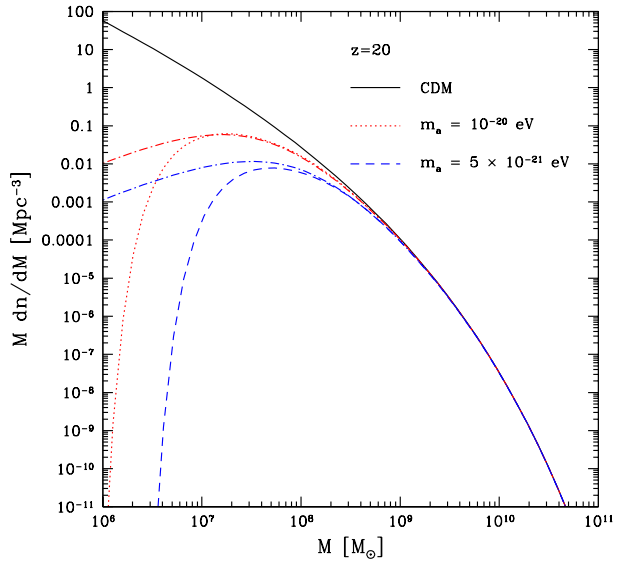}
\caption{The model halo mass functions at $z=20$ in CDM and FDM. The mass function in the FDM models is reduced relative to CDM below a mass scale of $\sim M_{1/2}$  (see Eq.~\ref{eq:mhalf}).
The dotted/dashed lines show the Marsh \& Silk \citep{Marsh:2013ywa} mass function, while the dot-dashed lines show results from the fitting formula of Schive et al. \citep{Schive:2015kza}. 
For the FDM masses shown, the two models agree fairly well near the suppression mass, and differ only on small mass scales (see text: in the Marsh \& Silk model the mass function is truncated below $\sim 0.01 M_{1/2}$), which have little impact on the overall collapse fraction and our $z \sim 20$ Ly-$\alpha$ background calculations.
In each model the FDM mass function is suppressed relative to CDM; this reduces the $z \sim 20$ Ly-$\alpha$ background and allows us to constrain $m_a$.}
\label{fig:fdm_massfunc}
\ec
\end{figure}
Fig.~\ref{fig:fdm_massfunc} shows an illustrative example, contrasting the $z=20$ mass function in CDM and FDM for two representative choices of axion mass, $m_a = 5 \times 10^{-21} {\rm eV}$
and $m_a = 10^{-20} {\rm eV}$.  Reassuringly, the models of Marsh \& Silk \citep{Marsh:2013ywa} and Schive et al. \citep{Schive:2015kza} agree fairly well near the suppression mass scale (at least for these values of $m_a$ and $z=20$), and differ only at relatively small halo mass. In practice, the halo mass function is so suppressed at small mass scales that our results on the $z \sim 20$ Ly-$\alpha$ background are relatively insensitive to which model we assume. 
In practice, we adopt the Marsh \& Silk \citep{Marsh:2013ywa} mass function in what follows, but comment explicitly on how the results differ if we instead assume the \citep{Schive:2015kza} fitting formula.  
In summary, the absence of small mass halos in FDM should allow us to constrain the axion mass using the global 21 cm signal. 

\subsection{Star Formation Efficiency and Minimum Halo Mass}
\label{sec:sfr_param}

Finally, we briefly discuss our model choices for the star formation efficiency and minimum host halo mass. One plausible value for the minimum halo mass hosting star formation is set by the mass at which the halo virial temperature reaches $T_{\rm vir} = 10^4 K$, above which atomic line cooling is efficient, allowing the gas to cool, condense, and form stars. If the gas in such halos is of primordial composition, and consists of highly ionized hydrogen/singly-ionized helium, then the mean molecular weight is $\mu=0.61$, and the total halo mass of a $T_{\rm vir} = 10^4$ K  halo at $z=20$ (in our assumed cosmology) is
$M=3.0 \times 10^7 M_\odot$ \cite{Barkana:2000fd}. It will also be helpful to note that the mass of a $T_{\rm vir} = 10^4$ K halo at $z=8$ is $M=1.1 \times 10^8 M_\odot$. Molecular hydrogen cooling may allow star formation in smaller mass halos, although this cooling channel will be suppressed as early stars turn on and emit dissociating UV radiation \cite{Haiman:1996rc}. For reference, the lowest minimum mass we consider, $M=10^6 M_\odot$, corresponds to a virial temperature of $T_{\rm vir} = 2,000$ K at $z=20$ (assuming primordial neutral gas and $\mu=1.22$ for these lower mass halos). In practice, we vary the minimum host halo mass across the broad range of $M_{\rm min}=10^6 - 10^9 M_\odot$. Our constraints on FDM, however, are insensitive to the choice of $M_{\rm min}$ provided that it is small compared to the FDM suppression mass scale, $M_{1/2}$. 

More important for our purposes is to consider a reasonable range of star formation efficiency parameters. This is obviously quite uncertain, as we don't have other observations at the high 
redshifts ($z \sim 20$) and low halo
masses ($M \sim 10^7-10^8 M_\odot$) of interest. Nevertheless, one handle on the star formation efficiency parameter in our models comes from comparing their predictions of the co-moving star-formation rate density (SFRD, or $\dot{\rho}_\star(z)$) with observed values (at the highest redshifts available), as inferred from measurements of UV luminosity functions. 
The SFRD in our model may be calculated as (see Eq.~\ref{eq:emiss}):
\beqa
\dot{\rho}_\star(z) = f_\star n_H(z=0) \frac{df_{\rm coll}(> M_{\rm min}, z)}{dt}.
\label{eq:sfrd}
\eeqa
We can then compare with the observed SFRD inferred from dust-corrected Schechter fits \cite{Schechter76} to the star formation rate functions (which describe the co-moving abundance
of galaxies of varying star formation rate) as derived using UV luminosity function measurements \cite{Mashian16}. At $z=7.9$ if one includes only star formation in galaxies above the current UV luminosity function detection limits of $M_{\rm UV} = -17.7$ mag, $\dot{\rho}_\star(z=7.9) = 6.8 \times 10^{-3} M_\odot {\rm yr}^{-1} {\rm Mpc}^{-3}$. On the other hand, if one integrates all the way down the Schechter function to zero star formation rate, the best-fit Schechter function parameters from \cite{Mashian16} give $\dot{\rho}_\star(z=7.9) = 0.014 M_\odot {\rm yr}^{-1} {\rm Mpc}^{-3}$. In our CDM models, this result (i.e. the SFRD extrapolated down the Schechter function to zero star formation rate) implies that $f_\star \sim 0.01$ if the minimum host mass is set by the atomic cooling limit, $M_{\rm min} = 1.1 \times 10^8 M_\odot$. In the FDM models of interest here with $m_a \geq 5 \times 10^{-21} {\rm eV}$, we find that essentially the same star formation efficiencies are required to
match the $z=8$ SFRD as in the CDM case. This results because the SFRD at $z \sim 8$ is dominated by star formation in higher mass halos than at $z \sim 20$ and because the truncation in the Marsh \& Silk \cite{Marsh:2013ywa} halo mass function moves to lower halo mass at decreasing redshift. 

We can also compare with abundance-matching constraints on the star-formation efficiency from the literature (\cite{Behroozi:2014tna,Sun16}). These studies favor a low star-formation efficiency in small mass halos: they find peak efficiencies of $f_\star \sim 0.2-0.3$ in halos with mass between  $M_{\rm halo} \sim 10^{11}-10^{12} M_\odot$ and steep declines towards smaller masses, with efficiencies falling below $f_\star \lesssim 0.01$ for $M_{\rm halo} \lesssim 10^{10} M_\odot$ at $z=8$ (see e.g. Fig. 2 of \cite{Sun16}). 
Note, however, that these constraints are limited
to halos of $M_{\rm halo} \gtrsim 10^{10} M_\odot$ at $z \sim 8$ and so large extrapolations are required to reach the atomic cooling mass and to move toward higher redshifts.
The steep decline toward small mass is usually attributed to supernova feedback, which may be less effective in compact galaxies at high redshift (e.g. \cite{Sun16}).  

A final potential handle on our model parameters comes from the electron scattering optical depth to CMB photons, $\tau_e$, although the utility of this is limited -- for our present purposes -- by the significant uncertainties in the escape fraction of ionizing photons, among other quantities. Nevertheless, we find that a model with $f_\star=0.01$, $M_{\rm min} = 3 \times 10^7 M_\odot$, produces a reasonable optical depth of 
$\tau_e = 0.07$ provided the escape fraction of ionizing photons is $f_{\rm esc}=0.2$, and  assuming a clumping factor of $C=2$, $N_{\rm ion} = 5,000$ ionizing
photons per stellar baryon  (see e.g. \cite{Lidz:2015ewe} for a discussion of these parameters.)  This is consistent with constraints from Planck data \cite{Aghanim:2015xee,Adam:2016hgk,Heinrich:2016ojb}. In this context, we note in passing that the principle component analysis of Heinrich et al. \cite{Heinrich:2016ojb} finds a hint for a contribution to the optical depth from very high redshift ($z \gtrsim 15$) using the Planck 2015 Low-Frequency Instrument (LFI) data. 
This might provide an independent avenue for constraining FDM, although recent work bounds the contribution to $\tau_e$ from high redshift using Planck High Frequency Instrument (HFI) data \citep{Millea:2018bko}. It will be interesting to see whether the (forthcoming) final release of CMB polarization data from the Planck collaboration shows evidence for high redshift contributions to the optical depth.

In light of these uncertainties, we adopt a simple halo mass-independent star formation efficiency here and assume that $f_\star \leq 0.05$ for the halo masses and redshifts of interest. We believe this is a conservative assumption: if we extrapolate our models to $z \sim 8$ with this efficiency, we overproduce the observed SFRD by a factor of several.
Nevertheless, it is worth emphasizing that one can circumvent the $z \sim 8$ star formation constraint by postulating a lower $f_\star$ at $z \sim 8$ compared to $z \sim 20$,
or a lower $f_\star$ for the more massive halos relevant for the $z \sim 8$ observations.
In any case, our model Ly-$\alpha$ specific intensity (Eq.~\ref{eq:jalpha}) is simply proportional to the star-formation efficiency, and so the reader can rescale our results as they see 
fit (see also Fig.~\ref{fig:zcouple_fdm}).

\section{Results} 

\begin{figure}[htpb]
\bc
\includegraphics[width=1.0\columnwidth]{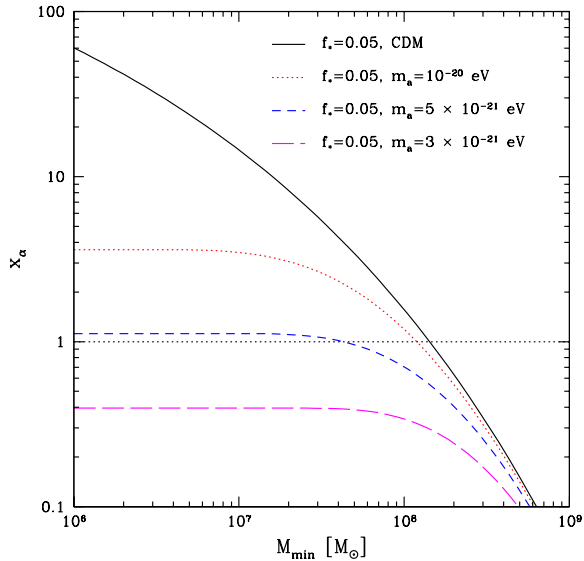}
\caption{The $z=20$ coupling coefficient, $x_\alpha$, in CDM and FDM. The curves show the coupling coefficient as a function of the minimum halo mass hosting star formation, assuming a halo mass independent star formation efficiency of $f_\star=0.05$. 
The EDGES result suggests that $x_\alpha \gtrsim 1$ at $z=20$ and so the FDM model with
$m_a = 5 \times 10^{-21} {\rm eV}$ just barely satisfies this constraint, and only does so provided stars form in sufficiently low mass halos (near the atomic cooling mass, see text) with the optimistic efficiency factor of $f_\star = 0.05$ considered here. Lower FDM particle masses are disfavored by the EDGES result. 
}
\label{fig:fdm_xalpha}
\ec
\end{figure}

We can now piece together the ingredients of our model and predict the Ly-$\alpha$ background at $z=20$ in CDM and FDM.  Specifically, we predict the coupling coefficient $x_\alpha$ (Eq.~\ref{eq:coupling_co}) in CDM and the two representative FDM models of Fig.~\ref{fig:fdm_massfunc} for $f_\star=0.05$ as a function of $M_{\rm min}$ using Eqs.~\ref{eq:jalpha}-\ref{eq:emiss}. We
also examine a third case, with slightly lower FDM mass, $m_a = 3 \times 10^{-21} {\rm eV}$. 
The results of these calculations are shown in Fig.~\ref{fig:fdm_xalpha}.
First, we note that even the CDM models with $f_\star = 0.05$ fail to achieve coupling at $z=20$ if $M_{\min} \gtrsim 2 \times 10^8 M_\odot$ and so the EDGES results suggests fairly efficient
star formation in low mass halos (see also \citep{Mirocha:2018cih}). Since the collapse fraction is a steep function of $M_{\rm min}$ in CDM, the star formation efficiency required to achieve $x_{\alpha}=1$ at $z=20$ goes down
significantly as $M_{\rm min}$ decreases. For example, at the atomic cooling mass limit of $M_{\rm min} = 3.0 \times 10^7 M_\odot$, the required star-formation efficiency is $f_\star \sim 0.01$.
Interestingly, this is the efficiency required to match the observed (Schechter-function extrapolated) SFRD at $z=8$ in our model (as discussed in the previous section), suggesting that the onset redshift is plausible in CDM models. 

In FDM, the suppressed halo mass function leads to smaller values of the coupling coefficient $x_\alpha$ and the curves flatten for $M_{\rm min}$ a bit smaller than the suppression mass, $M_{1/2}$ (Eq.~\ref{eq:mhalf}). The model with $m_a = 10^{-20} {\rm eV}$ can achieve coupling provided the star-formation efficiency is a little larger than $f_\star \gtrsim 0.01$, while the model with $m_a = 5 \times 10^{-21} {\rm eV}$ just achieves $x_{\alpha} = 1$ at $z=20$ for $f_\star = 0.05$. Models with smaller axion particle mass, such as the $m_a = 3 \times 10^{-21} {\rm eV}$ case shown in the figure, fall short of producing $x_{\alpha}=1$ by $z=20$. In the canonical case that $m_a=10^{-22} {\rm eV}$, the coupling coefficient falls significantly below the lowest value shown on the y-axis in Fig.~\ref{fig:fdm_xalpha} and so these models are strongly disfavored by EDGES.
The FDM results in Fig.~\ref{fig:fdm_xalpha} assume the Marsh \& Silk halo mass function \citep{Marsh:2013ywa}. If we instead adopt the Schive et al. \citep{Schive:2015kza} fitting formula, the coupling coefficient $x_\alpha$ at small minimum halo mass is $22\%$ larger for $m_a = 5 \times 10^{-21} {\rm eV}$ and $6\%$ smaller for $m_a = 10^{-20} {\rm eV}$, while the  difference is a bit more significant for $m_a = 3 \times 10^{-21} {\rm eV}$, in which case the Schive et al. formula predicts a larger $x_\alpha$ by $56\%$.\footnote{The 
Marsh \& Silk \citep{Marsh:2013ywa} model gives a slightly larger mass function close to the suppression mass for $m_a=10^{-20} {\rm eV}$, which boosts $x_\alpha$ relative to the \citep{Schive:2015kza} case in spite of the extra suppression at $\sim 0.01 M_{1/2}$ in the Marsh \& Silk model.} These differences are small compared to other uncertainties in our modeling, such as the choice of
$f_\star$.

In summary, under the fiducial assumptions adopted in this paper, the EDGES measurement disfavors models with $m_a \leq 5 \times 10^{-21} {\rm eV}$.
These results can also be roughly recast to constrain warm dark matter (WDM) models. Specifically, assuming that the WDM particles are thermal relics, we can match the suppression scale $k_{1/2}$
in WDM and FDM (see Eq.~\ref{eq:mhalf} and Eq. 63 of \cite{Hui:2016ltb}) for our limiting FDM particle mass of $m_a = 5 \times 10^{-21} {\rm eV}$. This gives $m_{\rm WDM} \geq 5 {\rm keV}$. 
These results are consistent with the other independent studies mentioned in the Introduction: \cite{Safarzadeh:2018hhg} places a bound on thermal relic WDM of $m_{\rm WDM} \geq 4 {\rm keV}$,
while \cite{Schneider:2018xba}'s constraint on FDM is $m_a \geq 8 \times 10^{-21} {\rm eV}$. The small differences with our limits likely reflect slightly different modeling choices regarding the star formation efficiency, onset redshift, and other parameters. 

\begin{figure}[htpb]
\bc
\includegraphics[width=1.0\columnwidth]{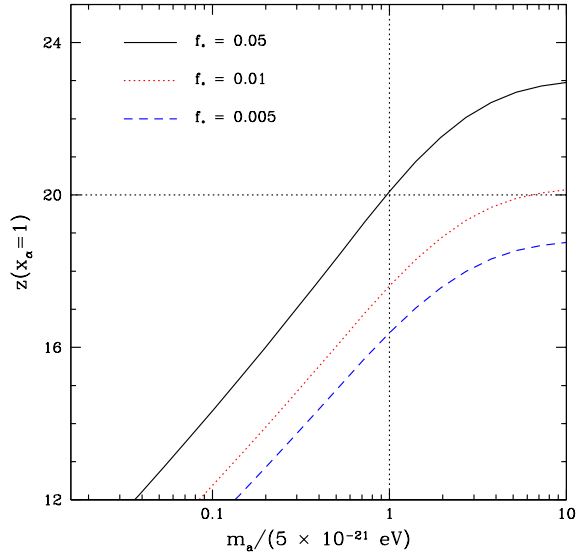}
\caption{The redshift at which the coupling constant, $x_\alpha$, first reaches unity for different FDM particle masses and star formation efficiencies. In each case, $M_{\rm min}$ is set
to the atomic cooling mass at $z=20$, $M = 3.0 \times 10^7 M_\odot$. The dotted lines indicate the coupling redshift suggested by EDGES and the implied lower bound on the axion mass (for $f_\star \leq 0.05)$.
}
\label{fig:zcouple_fdm}
\ec
\end{figure}

As acknowledged in the EDGES paper itself \cite{Bowman:2018yin}, an independent confirmation of the measurement is important, especially given the systematic challenges involved in the global redshifted 21 cm observations. We therefore explore how the redshift at which the coupling coefficient reaches unity ($x_{\alpha} = 1$) depends on the FDM particle mass and $f_\star$. These results will be useful in case the EDGES results are revised. This investigation also serves to test the sensitivity to our uncertain assumption that EDGES implies  $x_{\alpha}=1$ at $z=20$.  

The results of these calculations are shown in Fig.~\ref{fig:zcouple_fdm} for the case of $M_{\rm min} = 3.0 \times 10^7 M_\odot$ (the atomic cooling mass at $z=20$).  The curves also show explicitly how the coupling redshift depends on the star formation efficiency. For example, if the star formation efficiency is as low as $1\%$, achieving coupling by $z=20$ requires $m_a \gtrsim 2.5 \times 10^{-20} {\rm eV}$. One can also see how the constraints change depending on when coupling is achieved. For example, if the redshift where $x_{\alpha}$ first equals unity is revised downwards to $z(x_\alpha=1)=15$, then the constraint on FDM mass implied is $m_a \geq 6.6 \times 10^{-22} {\rm eV}$ (provided $f_\star \leq 0.05$). For the reader interested in thermal relic WDM, a rough translation from the FDM bound can be made as: $m_{\rm WDM} = 5 {\rm keV} \left[m_a/(5 \times 10^{-21} {\rm eV}) \right]^{0.4}$ (e.g. \cite{Hui:2016ltb}).

\section{Conclusions}

We have explored how the onset redshift of the 21 cm absorption signal in the globally averaged redshifted 21 cm signal may be used to constrain the FDM particle mass. 
Taken at face value, the recent EDGES measurement implies an interesting mass limit of $m_a \geq 5 \times 10^{-21} {\rm eV}$ in our models. In the case of thermal relic WDM, this limit
translates roughly to $m_{\rm WDM} \geq 5 {\rm keV}$, by the method of matching the suppression scale in the initial power spectrum (see e.g. \cite{Hui:2016ltb} for a discussion).
The FDM constraints are comparable or slightly tighter than those in the current literature, with the sharpest present limits coming from Ly-$\alpha$ forest observations: for example, \cite{Irsic:2017yje} find $m_a \geq 2 \times 10^{-21} {\rm eV}$, while an independent analysis from \cite{Armengaud:2017nkf} places a limit of $m_a \geq 2.9 \times 10^{-21} {\rm eV}$ (both results are at the $2-\sigma$ confidence level). 
The Ly-$\alpha$ forest constraint is predicated upon the correct modeling of astrophysical fluctuations such as in the ionizing background, in the temperature and from feedback processes
(a discussion can be found in \cite{Hui:2016ltb}). 
The 21 cm constraint presented in this paper has its own assumptions and caveats as well:
 the results depend on modeling the halo mass function in FDM and should be refined using
future FDM simulations; the constraints assume a star-formation efficiency of $f_\star \leq 0.05$ for the halo masses/redshifts of interest; and finally
we assume that $x_\alpha=1$ at $z=20$. As the measurements improve, a more detailed comparison between the models and observations will be warranted: this would include, for example, a chi-squared analysis using the full observed absorption profile, incorporating statistical and systematic error bars, and marginalizing over nuisance parameters in the model parameter space. In principle,  fitting the full global signal in conjunction with fluctuation measurements should help break degeneracies with the uncertain star formation efficiency parameter, and its possible redshift evolution \cite{Sitwell:2013fpa}. 

In the near term, other global redshifted 21 cm experiments such as those described in \cite{Singh:2017syr,Price17,Peterson:2014rga} are poised to confirm or refute the EDGES results. In addition, the HERA project \cite{DeBoer:2016tnn} has the bandwidth and sensitivity to detect fluctuations in the 21 cm brightness temperature from the Cosmic Dawn era \cite{Mesinger:2013nua}, especially if the absorption signal is as pronounced (on average) as implied by the EDGES measurements.  It will be interesting to explore the implications of these upcoming measurements for FDM.

\acknowledgements

We thank Angus Beane, Tzu-Ching Chang, Olivier Dor\'e, Aaron Ewall-Wice, Colin Hill, Jerry Ostriker, Jacqueline van Gorkom, and Eli Visbal for numerous and inspiring discussions about EDGES. This research is supported in part by NASA grant NXX16AB27G and DOE grant DE-SC0011941.

\bibliographystyle{hieeetr}

\bibliography{references}

\begin{thebibliography}{10}

\bibitem{Furlanetto:2006jb}
S.~Furlanetto, S.~P. Oh, and F.~Briggs, ``{Cosmology at Low Frequencies: The 21
  cm Transition and the High-Redshift Universe},'' {\em Phys. Rept.}, vol.~433,
  pp.~181--301, 2006, astro-ph/0608032.

\bibitem{Pritchard2012}
J.~R. {Pritchard} and A.~{Loeb}, ``{21 cm cosmology in the 21st century},''
  {\em Reports on Progress in Physics}, vol.~75, p.~086901, Aug. 2012,
  1109.6012.

\bibitem{Bowman:2018yin}
J.~D. Bowman, A.~E.~E. Rogers, R.~A. Monsalve, T.~J. Mozdzen, and N.~Mahesh,
  ``{An absorption profile centred at 78 megahertz in the sky-averaged
  spectrum},'' {\em Nature}, vol.~555, no.~7694, pp.~67--70, 2018.

\bibitem{Barkana:2018lgd}
R.~Barkana, ``{Possible interaction between baryons and dark-matter particles
  revealed by the first stars},'' {\em Nature}, vol.~555, no.~7694, pp.~71--74,
  2018, 1803.06698.

\bibitem{Munoz:2018pzp}
J.~B. Muñoz and A.~Loeb, ``{Insights on Dark Matter from Hydrogen during Cosmic
  Dawn},'' 2018, 1802.10094.

\bibitem{Berlin:2018sjs}
A.~Berlin, D.~Hooper, G.~Krnjaic, and S.~D. McDermott, ``{Severely Constraining
  Dark Matter Interpretations of the 21-cm Anomaly},'' 2018, 1803.02804.

\bibitem{Ewall-Wice:2018bzf}
A.~Ewall-Wice, T.~C. Chang, J.~Lazio, O.~Dore, M.~Seiffert, and R.~A. Monsalve,
  ``{Modeling the Radio Background from the First Black Holes at Cosmic Dawn:
  Implications for the 21 cm Absorption Amplitude},'' 2018, 1803.01815.

\bibitem{Mirocha:2018cih}
J.~Mirocha and S.~R. Furlanetto, ``{What does the first highly-redshifted 21-cm
  detection tell us about early galaxies?},'' 2018, 1803.03272.

\bibitem{Hill:2018lfx}
J.~C. Hill and E.~J. Baxter, ``{Can Early Dark Energy Explain EDGES?},'' 2018,
  1803.07555.

\bibitem{Venumadhav:2018uwn}
T.~Venumadhav, L.~Dai, A.~Kaurov, and M.~Zaldarriaga, ``{Heating of the
  intergalactic medium by the cosmic microwave background during cosmic
  dawn},'' 2018, 1804.02406.

\bibitem{Munoz:2015bca}
J.~B. Muñoz, E.~D. Kovetz, and Y.~Ali-Haïmoud, ``{Heating of Baryons due to
  Scattering with Dark Matter During the Dark Ages},'' {\em Phys. Rev.},
  vol.~D92, no.~8, p.~083528, 2015, 1509.00029.

\bibitem{Feng:2018rje}
C.~Feng and G.~Holder, ``{Enhanced global signal of neutral hydrogen due to
  excess radiation at cosmic dawn},'' 2018, 1802.07432.

\bibitem{Sharma:2018agu}
P.~Sharma, ``{Astrophysical radio background cannot explain the EDGES signal:
  constraints from cooling of non-thermal electrons},'' 2018, 1804.05843.

\bibitem{Wouthuysen52}
S.~A. {Wouthuysen}, ``{On the excitation mechanism of the 21-cm
  (radio-frequency) interstellar hydrogen emission line.},'' {\em \aj},
  vol.~57, pp.~31--32, 1952.

\bibitem{Field58}
G.~B. {Field}, ``{Excitation of the Hydrogen 21-CM Line},'' {\em Proceedings of
  the IRE}, vol.~46, pp.~240--250, Jan. 1958.

\bibitem{Hirata:2005mz}
C.~M. Hirata, ``{Wouthuysen-Field coupling strength and application to
  high-redshift 21 cm radiation},'' {\em Mon. Not. Roy. Astron. Soc.},
  vol.~367, pp.~259--274, 2006, astro-ph/0507102.

\bibitem{Pritchard:2005an}
J.~R. Pritchard and S.~R. Furlanetto, ``{Descending from on high: lyman series
  cascades and spin-kinetic temperature coupling in the 21 cm line},'' {\em
  Mon. Not. Roy. Astron. Soc.}, vol.~367, pp.~1057--1066, 2006,
  astro-ph/0508381.

\bibitem{Gunn:1965hd}
J.~E. Gunn and B.~A. Peterson, ``{On the Density of Neutral Hydrogen in
  Intergalactic Space},'' {\em Astrophys. J.}, vol.~142, p.~1633, 1965.

\bibitem{Chen:2003gc}
X.-L. Chen and J.~Miralda-Escude, ``{The spin - kinetic temperature coupling
  and the heating rate due to Lyman - alpha scattering before reionization:
  Predictions for 21cm emission and absorption},'' {\em Astrophys. J.},
  vol.~602, pp.~1--11, 2004, astro-ph/0303395.

\bibitem{Hu:2000ke}
W.~Hu, R.~Barkana, and A.~Gruzinov, ``{Cold and fuzzy dark matter},'' {\em
  Phys. Rev. Lett.}, vol.~85, pp.~1158--1161, 2000, astro-ph/0003365.

\bibitem{Hui:2016ltb}
L.~Hui, J.~P. Ostriker, S.~Tremaine, and E.~Witten, ``{Ultralight scalars as
  cosmological dark matter},'' {\em Phys. Rev.}, vol.~D95, no.~4, p.~043541,
  2017, 1610.08297.

\bibitem{Bozek:2014uqa}
B.~Bozek, D.~J.~E. Marsh, J.~Silk, and R.~F.~G. Wyse, ``{Galaxy UV-luminosity
  function and reionization constraints on axion dark matter},'' {\em Mon. Not.
  Roy. Astron. Soc.}, vol.~450, no.~1, pp.~209--222, 2015, 1409.3544.

\bibitem{Barkana:2001gr}
R.~Barkana, Z.~Haiman, and J.~P. Ostriker, ``{Constraints on warm dark matter
  from cosmological reionization},'' {\em Astrophys. J.}, vol.~558, p.~482,
  2001, astro-ph/0102304.

\bibitem{Sitwell:2013fpa}
M.~Sitwell, A.~Mesinger, Y.-Z. Ma, and K.~Sigurdson, ``{The Imprint of Warm
  Dark Matter on the Cosmological 21-cm Signal},'' {\em Mon. Not. Roy. Astron.
  Soc.}, vol.~438, no.~3, pp.~2664--2671, 2014, 1310.0029.

\bibitem{Mesinger:2013nua}
A.~Mesinger, A.~Ewall-Wice, and J.~Hewitt, ``{Reionization and beyond:
  detecting the peaks of the cosmological 21?cm signal},'' {\em Mon. Not. Roy.
  Astron. Soc.}, vol.~439, no.~4, pp.~3262--3274, 2014, 1310.0465.

\bibitem{Safarzadeh:2018hhg}
M.~Safarzadeh, E.~Scannapieco, and A.~Babul, ``{A limit on the warm dark matter
  particle mass from the redshifted 21 cm absorption line},'' 2018, 1803.08039.

\bibitem{Schneider:2018xba}
A.~Schneider, ``{Constraining Non-Cold Dark Matter Models with the Global 21-cm
  Signal},'' 2018, 1805.00021.

\bibitem{Ade:2015xua}
P.~A.~R. Ade {\em et~al.}, ``{Planck 2015 results. XIII. Cosmological
  parameters},'' {\em Astron. Astrophys.}, vol.~594, p.~A13, 2016, 1502.01589.

\bibitem{Eisenstein:1997jh}
D.~J. Eisenstein and W.~Hu, ``{Power spectra for cold dark matter and its
  variants},'' {\em Astrophys. J.}, vol.~511, p.~5, 1997, astro-ph/9710252.

\bibitem{Madau:1996cs}
P.~Madau, A.~Meiksin, and M.~J. Rees, ``{21-CM tomography of the intergalactic
  medium at high redshift},'' {\em Astrophys. J.}, vol.~475, p.~429, 1997,
  astro-ph/9608010.

\bibitem{Furlanetto:2006fs}
S.~Furlanetto and J.~R. Pritchard, ``{The Scattering of Lyman-series Photons in
  the Intergalactic Medium},'' {\em Mon. Not. Roy. Astron. Soc.}, vol.~372,
  pp.~1093--1103, 2006, astro-ph/0605680.

\bibitem{Chuzhoy:2006au}
L.~Chuzhoy and P.~R. Shapiro, ``{Heating and cooling of the intergalactic
  medium by resonance photons},'' {\em Astrophys. J.}, vol.~655, pp.~843--846,
  2007, astro-ph/0604483.

\bibitem{Barkana:2004vb}
R.~Barkana and A.~Loeb, ``{Detecting the earliest galaxies through two new
  sources of 21cm fluctuations},'' {\em Astrophys. J.}, vol.~626, pp.~1--11,
  2005, astro-ph/0410129.

\bibitem{Furlanetto:2006tf}
S.~Furlanetto, ``{The Global 21 Centimeter Background from High Redshifts},''
  {\em Mon. Not. Roy. Astron. Soc.}, vol.~371, pp.~867--878, 2006,
  astro-ph/0604040.

\bibitem{Tseliakhovich10}
D.~{Tseliakhovich} and C.~{Hirata}, ``{Relative velocity of dark matter and
  baryonic fluids and the formation of the first structures},'' {\em \prd},
  vol.~82, p.~083520, Oct. 2010, 1005.2416.

\bibitem{Fialkov:2013jxh}
A.~Fialkov, R.~Barkana, A.~Pinhas, and E.~Visbal, ``{Complete history of the
  observable 21-cm signal from the first stars during the pre-reionization
  era},'' {\em Mon. Not. Roy. Astron. Soc.}, vol.~437, p.~36, 2014, 1306.2354.

\bibitem{Schive:2014dra}
H.-Y. Schive, T.~Chiueh, and T.~Broadhurst, ``{Cosmic Structure as the Quantum
  Interference of a Coherent Dark Wave},'' {\em Nature Phys.}, vol.~10,
  pp.~496--499, 2014, 1406.6586.

\bibitem{Mocz:2017wlg}
P.~Mocz, M.~Vogelsberger, V.~H. Robles, J.~Zavala, M.~Boylan-Kolchin,
  A.~Fialkov, and L.~Hernquist, ``{Galaxy formation with BECDM ? I. Turbulence
  and relaxation of idealized haloes},'' {\em Mon. Not. Roy. Astron. Soc.},
  vol.~471, no.~4, pp.~4559--4570, 2017, 1705.05845.

\bibitem{Veltmaat:2018dfz}
J.~Veltmaat, J.~C. Niemeyer, and B.~Schwabe, ``{Formation and structure of
  ultralight bosonic dark matter halos},'' 2018, 1804.09647.

\bibitem{Schive:2015kza}
H.-Y. Schive, T.~Chiueh, T.~Broadhurst, and K.-W. Huang, ``{Contrasting Galaxy
  Formation from Quantum Wave Dark Matter, $\psi$DM, with $\Lambda$CDM, using
  Planck and Hubble Data},'' {\em Astrophys. J.}, vol.~818, no.~1, p.~89, 2016,
  1508.04621.

\bibitem{Sheth:1999mn}
R.~K. Sheth and G.~Tormen, ``{Large scale bias and the peak background
  split},'' {\em Mon. Not. Roy. Astron. Soc.}, vol.~308, p.~119, 1999,
  astro-ph/9901122.

\bibitem{Marsh:2013ywa}
D.~J.~E. Marsh and J.~Silk, ``{A Model For Halo Formation With Axion Mixed Dark
  Matter},'' {\em Mon. Not. Roy. Astron. Soc.}, vol.~437, no.~3,
  pp.~2652--2663, 2014, 1307.1705.

\bibitem{Marsh:2016vgj}
D.~J.~E. Marsh, ``{WarmAndFuzzy: the halo model beyond CDM},'' 2016,
  1605.05973.

\bibitem{Barkana:2000fd}
R.~Barkana and A.~Loeb, ``{In the beginning: The First sources of light and the
  reionization of the Universe},'' {\em Phys. Rept.}, vol.~349, pp.~125--238,
  2001, astro-ph/0010468.

\bibitem{Haiman:1996rc}
Z.~Haiman, M.~J. Rees, and A.~Loeb, ``{Destruction of molecular hydrogen during
  cosmological reionization},'' {\em Astrophys. J.}, vol.~476, p.~458, 1997,
  astro-ph/9608130.

\bibitem{Schechter76}
P.~{Schechter}, ``{An analytic expression for the luminosity function for
  galaxies.},'' {\em \apj}, vol.~203, pp.~297--306, Jan. 1976.

\bibitem{Mashian16}
N.~{Mashian}, P.~A. {Oesch}, and A.~{Loeb}, ``{An empirical model for the
  galaxy luminosity and star formation rate function at high redshift},'' {\em
  \mnras}, vol.~455, pp.~2101--2109, Jan. 2016, 1507.00999.

\bibitem{Behroozi:2014tna}
P.~S. Behroozi and J.~Silk, ``{A Simple Technique for Predicting High-Redshift
  Galaxy Evolution},'' {\em Astrophys. J.}, vol.~799, no.~1, p.~32, 2015,
  1404.5299.

\bibitem{Sun16}
G.~{Sun} and S.~R. {Furlanetto}, ``{Constraints on the star formation
  efficiency of galaxies during the epoch of reionization},'' {\em \mnras},
  vol.~460, pp.~417--433, July 2016, 1512.06219.

\bibitem{Lidz:2015ewe}
A.~Lidz, ``{Modeling the Intergalactic Medium during the Epoch of
  Reionization},'' 2015, 1511.01188.

\bibitem{Aghanim:2015xee}
N.~Aghanim {\em et~al.}, ``{Planck 2015 results. XI. CMB power spectra,
  likelihoods, and robustness of parameters},'' {\em Astron. Astrophys.},
  vol.~594, p.~A11, 2016, 1507.02704.

\bibitem{Adam:2016hgk}
R.~Adam {\em et~al.}, ``{Planck intermediate results. XLVII. Planck constraints
  on reionization history},'' {\em Astron. Astrophys.}, vol.~596, p.~A108,
  2016, 1605.03507.

\bibitem{Heinrich:2016ojb}
C.~H. Heinrich, V.~Miranda, and W.~Hu, ``{Complete Reionization Constraints
  from Planck 2015 Polarization},'' {\em Phys. Rev.}, vol.~D95, no.~2,
  p.~023513, 2017, 1609.04788.

\bibitem{Millea:2018bko}
M.~Millea and F.~Bouchet, ``{Cosmic Microwave Background Constraints in Light
  of Priors Over Reionization Histories},'' 2018, 1804.08476.

\bibitem{Irsic:2017yje}
V.~Ir¨i?, M.~Viel, M.~G. Haehnelt, J.~S. Bolton, and G.~D. Becker, ``{First
  constraints on fuzzy dark matter from Lyman-$\alpha$ forest data and
  hydrodynamical simulations},'' {\em Phys. Rev. Lett.}, vol.~119, no.~3,
  p.~031302, 2017, 1703.04683.

\bibitem{Armengaud:2017nkf}
E.~Armengaud, N.~Palanque-Delabrouille, C.~Yèche, D.~J.~E. Marsh, and J.~Baur,
  ``{Constraining the mass of light bosonic dark matter using SDSS
  Lyman-$\alpha$ forest},'' {\em Mon. Not. Roy. Astron. Soc.}, vol.~471, no.~4,
  pp.~4606--4614, 2017, 1703.09126.

\bibitem{Singh:2017syr}
S.~Singh, R.~Subrahmanyan, N.~U. Shankar, M.~S. Rao, B.~S. Girish,
  A.~Raghunathan, R.~Somashekar, and K.~S. Srivani, ``{SARAS 2: A Spectral
  Radiometer for probing Cosmic Dawn and the Epoch of Reionization through
  detection of the global 21 cm signal},'' 2017, 1710.01101.

\bibitem{Price17}
D.~C. {Price}, L.~J. {Greenhill}, A.~{Fialkov}, G.~{Bernardi}, H.~{Garsden},
  B.~R. {Barsdell}, J.~{Kocz}, M.~M. {Anderson}, S.~A. {Bourke}, J.~{Craig},
  M.~R. {Dexter}, J.~{Dowell}, M.~W. {Eastwood}, T.~{Eftekhari}, S.~W.
  {Ellingson}, G.~{Hallinan}, J.~M. {Hartman}, R.~{Kimberk}, T.~J.~W. {Lazio},
  S.~{Leiker}, D.~{MacMahon}, R.~{Monroe}, F.~{Schinzel}, G.~B. {Taylor},
  D.~{Werthimer}, and D.~P. {Woody}, ``{Design and characterization of the
  Large-Aperture Experiment to Detect the Dark Age (LEDA) radiometer
  systems},'' {\em ArXiv e-prints}, Sept. 2017, 1709.09313.

\bibitem{Peterson:2014rga}
J.~B. Peterson, T.~C. Voytek, A.~Natarajan, J.~M.~J. Garcia, and O.~Lopez-Cruz,
  ``{Measuring the 21 cm Global Brightness Temperature Spectrum During the Dark
  Ages with the SCI-HI Experiment},'' in {\em {Proceedings, 49th Rencontres de
  Moriond on Cosmology: La Thuile, Italy, March 15-22, 2014}}, pp.~129--134,
  2014, 1409.2774.

\bibitem{DeBoer:2016tnn}
D.~R. DeBoer {\em et~al.}, ``{Hydrogen Epoch of Reionization Array (HERA)},''
  {\em Publ. Astron. Soc. Pac.}, vol.~129, p.~045001, 2017, 1606.07473.

\end{thebibliography}

\end{document}